# Work Issues in Software Engineering

Jeremy Leipzig
CSC 510 Software Engineering
North Carolina State University
December 6, 2002

# Abstract

Using data from a web-based survey of software developers, the author attempts to determine root causes of "death march" projects and excessive work hours in the software industry in relation to company practices and management. Special emphasis is placed on the factor of business/technical supervisor background. An analysis of variance revealed significant differences between these supervisor groups with regard to a "Pointy-Haired Boss" (PHB) sentiment index. This difference, combined with correlations between the PHB index and the endpoints of project failure and use of software engineering practices, indicate some disparity in the suitability of business-background supervisors to manage software development projects compared with their technical-background counterparts. Other survey data points to improved project management skills as the biggest necessity for supervisors in the business-background group.

# Table of Contents

"Software sucks because users demand it to."
~ Nathan Myhrvold (former CTO of Microsoft)

Labor disgraces no man, but occasionally men disgrace labor.
~ Ulysses S. Grant

# Introduction

The use of software engineering principles is, as Roger Pressman puts it, "conventional wisdom."[RP00]  With so much entrenched knowledge, it is still something of a mystery why more companies have not adopted better practices. A recent study by Carnegie Mellon's Software Engineering Institute reported 38% of survey organizations were at CMM level 1.[SE01]  Pressman is gracious enough to attribute this resistance to "the difficulty of technology transition and the cultural change that accompanies it."

Pointing to the lack application of software engineering principles in industry is no longer useful without an examination of the underlying reasons, however unquantifiable, of why such valuable advice has not been heeded.[CC95]

Projects completed at one of these outmoded establishments are often the end product of a "death march."  The software industry borrowed this term from war history presumably to describe the horrors programmers experience when involved in overly ambitious and time-limited projects primarily characterized by long hours, usually forced by the unyielding fist of a tyrannical supervisor.

This purpose of this paper and its accompanying survey is to study the root causes and effects of these death marches, behaviors of the companies that sponsor them and the management that oversees them, from the perspective of people who work on them.

**Consequences**
Special attention will be given to the long hours spent on such projects, as this factor tends to be their most significant for the primary stakeholders, the developers, for health and family reasons.  When work spills over into night, circadian rhythms that govern our physiological systems can be disrupted.  In addition to disturbing normal sleep patterns and daytime performance, the symptoms of chronically pulling all-nighters resemble those found in shift-workers – gastrointestinal problems, cognitive deficits, and heart problems.[SR01][LL02]  Time away from home has detrimental effects on families and communities, putting extra unpaid housework loads on spouses, and decreasing the available time for civic and volunteer activities.[JJ01]

Companies suffer from death marches in the form of more software errors due to a strained testing schedule.  However, one commonly overlooked price companies pay for overly enthusiastic or outright dishonest delivery estimates is their reputation.  The software industry already has a notoriety in the business world for vaporware and may be on the verge of an crisis with regard to the legal liability of software reliability, another software engineering issue.[RC95]

**Death March Defined**
Author of "Death March" Edward Yourdon establishes the criteria for this dubious distinction as when the schedule, staff, or budget is half of what it should be or when the project requirements are twice what they would normally be [EY97]. As awful as that sounds, Yourdon and others are quick to point out the possibility of positive death marches projects, where the stakeholders are motivated to work insane hours on exciting new projects. The majority of start-ups operate this way with willing compliance from their employees. The flip side of these cases are the long hours spent on routine, unrewarding, or unexciting projects in companies that are well capable of implementing more mature processes.

Yourdon (Table 1) offer an excellent spectrum of reasons for death march projects, although his list might appear to conceal the origin of most "unplanned crisis" – poor or non-existent software engineering methodologies. These reasons provide a foundation for identifying underlying characteristics of the software industry, individual companies practices, and the personalities and behaviors of management that are most likely to lead employees into death marches.

| Table 1: Yourdon's Reasons for Death March Projects [EY97] |
| --- |
| Politics, politics, politics. |
| Naïve promises made by marketing, senior executives, naïve project managers, etc. |
| Naïve optimism of youth |
| The "start-up" mentality |
| The "Marine Corps" mentality |
| Intense competition caused by globalization of markets. |
| Intense competition caused by the appearance of new technologies. |
| Intense pressure caused by unexpected government regulations. |
| Unexpected and/or unplanned crises |

**The Industry**

Software development is a more project-oriented industry than many of those in other business sectors, which means its organizations are inherently more susceptible to the vagaries of poor project management. Although major projects like highway construction and lawsuits can spin out of control, several aspects of the software industry have made it particularly susceptible to spawning death march projects.

First, as a creative process, software development offers greater autonomy to its developers than other forms of engineering. This increased independence translates to more responsibility placed on developers to manage their own projects. This in turn makes the software development more susceptible to the most of common of human faults - procrastination.

Parkinson's law states that "work expands to fill the time available." Most software development jobs are salaried positions with few restrictions on "business" hours, which especially prone to filling up boundless amount of time with work. "Why do employees need to work so many hours? The key is that management does not view employee time as a scarce resource," writes former Microsoft programmer Adam Barr [AB00].

In conventional high-intensity professional jobs like those found in law, medicine, and business, there is an obvious financial payoff to putting in long hours. Although the pay of software developers is decent, the real lure for the preceding decade has been stock options.  These speculative perks constitute huge deterrents for developers to demand payment of overtime, one device that could have a huge impact on the workweek of developers [AN99]. Demanding overtime is incongruent with the idea of distributed ownership and responsibility that implies.  Overtime is also associated with low-level employees, so developers who consider themselves professionals are often unwilling to accept overtime for fear of sustaining a status hit.  Labor unions, which could also regulate work hours, have not yet taken hold in the software industry, perhaps due to the supply-demand inequities and the aforementioned stock option bonanza developers enjoyed during the 1990's.

Although software development is perhaps equally time-intensive as other forms of engineering, the commercialization of the industry brings with new costs in terms of development time.  Start-ups in the software industry are highly dependent on demos to fuel investment, partnership, and consumer interest.  Because many demos are shallow interfaces, the work that goes into them cannot be transferred to a final product, meaning large expanses of time are lost to fluff.

**The Companies**
As mentioned, death marches are not inherently bad.  By extension, not all companies that sponsor death marches are bad either.  Start-ups, with limited windows of opportunity in which to release new products, constitute good exceptions to the rule.  Young or adventurous programmers eager to challenge themselves in these environments join them presumably knowing the risk and sacrifice demanded of them.

Many bigger and more mature companies become entrenched in "death march as a way of life" because they feel it offers some sort of advantage over a more methodical process.[EY97]  While not the worst offender in this area, Microsoft deserves special attention because of its commercial success and influence, which have propelled it to become a role model for many organizations.  For better or worse, Microsoft has geared itself to a death march culture of long hours and deadline crises.  The process of finding programmers to tolerate this begins with an elaborate recruiting process with a special emphasis on finding individuals who "fit in" to the company.  Hiring college graduates with little exposure to other, perhaps slower companies, Microsoft can in a sense mold these grads into death marchers using a kind of boot camp atmosphere.  With its sports fields, shower facilities, and casual atmosphere Microsoft has attempted to engineer their campus as an extension of college, with the requisite of all-nighters included.  The intensive recruitment process and campus experience are also ways in which Microsoft strengthens its employee community and serve as a platform for espousing the "us vs. them" attitude that has become emblematic of the company.  Along with stock options, these processes are the primary ways in which Microsoft is able to operate continuously in death march mode while other employees at other companies lose morale.

The project scheduling process known as "signing up" serves as a particularly insidious means Microsoft has developed of goading bigger efforts from its developers [AB00]. During the planning phase of a software project, a supervisor known as the program manager decides who will complete which tasks, but leaves estimation of these tasks up to the individual programmers to determine.  Once a developer has submitted a schedule, that schedule becomes a de facto commitment on their part to meet the deadline.  The problem with this arrangement, as Burr points out, is that realistic schedules must anticipate changes to the spec, integration issues, major bugs, and all the other problems that typically delay delivery.  Being lowly developers without access to all the project details, those responsible for planning cannot often justify padding their schedules to compensate for these delays, and instead turn in overly optimistic estimates.[AB00]

**Management**
Management, specifically mid-level managers, bear considerable responsibility for averting death marches and keeping work hours reasonable.  One reason managers might fail to do this is a lack of incentive to apply software engineering principles that they feel are not helpful in the short term.  A better strategy from a manager's perspective might be to force a team to work a death march to get the project done, rather than worry about the long-term consequences on the people involved. "Extra time worked by employees is a tool that can be used to whitewash any number of bad decisions made by management" [AB00].  This sentiment appears in numerous web pages and message boards postings of agile methodology consultants.  Many of these experts claim management is more likely to provide lip service to their methodologies, supporting them only to the point they begin to conflict with the toxic operating procedures already in place.

**The Pointy-Haired Boss**
Although numerous articles have placed blame on a "lack of management support" [BC98], a cogent view of this bad manager is not provided.  Outside the academic literature, in American popular culture, the archetypal bad supervisor is clearly embodied by Dilbert's pointy-haired boss (PHB).

The PHB is myopic yet micro-manages, irrational, consummately political, unsympathetic, stupid, oppressive, takes credit for his employees' accomplishments. Unlike celebrity bosses whose technical competence (Gates), vision (Jobs), or charisma (Ellison) overcome any impersonal or confrontational manners, the PHB possesses none of these compensatory qualities.  The PHB is apathetic, ignorant, and is usually levels above where his technical and managerial competence ends.

> His top priorities are the bottom line and looking good in front of his subordinates and superiors (not necessarily in that order). Of absolutely no concern to him is the professional or personal well-being of his employees. The Boss is technologically challenged but he stays current on

> all the latest business trends, even though he rarely understands them.
>
> - from the Dilbert website:

The PHB is the cornerstone of Scott Adams' Dilbert Principle [SA96], which states that the "the most ineffective workers are systematically moved to the place where they can do the least damage: management." The Dilbert Principle is in essence a reformulation of the Peter Principle, which postulated that "in a hierarchy every employee tends to rise to his level of incompetence."[LA69]  Borrowing from these theories and the stereotypes they are derived from, we may infer that Adams and Peter feel the social and political qualities that enable an individual to climb high in the corporate ladder are uncorrelated or even *negatively correlated* with the crucial technical skills and intelligence needed to do the job.

**The Nerd Boss**
The nerd boss (NB) is a term used in this article to describe neophyte mangers that come from technical backgrounds.  Although software engineering literature does not use the term, extensive descriptions of this type have appeared Pressman [RP98] and Mackey [MK98] in the context of poor management.  While the NB can do the jobs of any of her subordinates and often works longer hours than them, she lacks people skills, is often politically inept, and reticent to delegate authority.  The NB might be just as likely to eschew software engineering principles because they entail doing more unpleasant management activities.  The NB more closely resembles the traditional "technical manager" stereotype as someone who is promoted by merit without possessing few managerial or social skills. Mackey claims these introverted technical types do not foster teamwork because "they frequently have a blind spot when it comes to human interactions." [MK98]

**Goals of the Survey**
One of the most interesting questions for the author is identifying which skills lacking in organizations lead to death march projects and who is not bringing these skills to the table.  From the programmer's perspective, the PHB lacks the technical skills required to realize an implementation of software engineering principles.[MW00]  Conversely, some experts feel the problem lies in the NB's lack of people skills needed to carry out the level of communication large projects require.[MK98]

The prime aims of the survey were to measure absolute frequency of NB and PHB's, as well as relation to death march factors such as excessive hours, communication problems, shortsightedness, and workplace conditions.  The survey was geared toward low-level software developers in companies with at least 3 people.

# Methods

A web-based questionnaire (Figure 1) of 29 items was developed (2 of these items were added midway into the polling, two others did not appear in the results until midway because of technical errors in the survey application). The majority of questions took the "sentiment-style" format popular with course evaluations. 8 were fill-in-the-blank.

Solicitations for survey takers were posted to various computer programming newsgroups and websites.

Results were analyzed using Graphpad Instat 3 for Macintosh.

4 Part-timers (those whose normal work week was <30 hrs) were removed from the data set.

A correlation matrix of quantifiable items was created by running Spearman non-parametric correlation statistics corrected for ties on every combination of items. P values are two-tailed.

Categorical questions were used to create quasi-experimental groups for use in one-way analysis of variance (ANOVA). Tukey post-hoc tests were used to determine which groups were significantly different.

Background and skill formed a matrix that was evaluated by chi-square analysis.

**Figure 1: Web survey**
http://zigster.com/survey.html

# Results

## Group statistics
Personal Info

|  | Age | # supervise | Using SE practices | Using agile methods | % failure |
|---|---|---|---|---|---|
| **Mean** | 33.06 | 1.22 | 3.24 | 0.16 | 10.57 |
| **Standard Deviation** | 9.95 | 3.51 | 1.28 | 0.37 | 18.14 |
| **Sample Size** | 116.00 | 115.00 | 117.00 | 111.00 | 104.00 |
| **Std. Error of mean** | 0.92 | 0.33 | 0.12 | 0.04 | 1.78 |
| **Minimum** | 17 | 0 | 1 | 0 | 0 |
| **Median** | 29.5 | 0 | 4 | 0 | 0 |
| **Maximum** | 60 | 30 | 5 | 1 | 100 |

| **Most likely to do in a crisis** | # | % |
|---|---|---|
| cut | 37 | 31.36 |
| extend | 66 | 55.93 |
| bust | 7 | 5.932 |
| dk | 5 | 4.237 |
| n/a | 3 | 2.542 |
| total | 118.00 | 100 |

The Workweek

|  | Avg. Hrs. | Max Hrs. | Boss forces long hours | Peer pressure | Health & Family |
|---|---|---|---|---|---|
| **Mean** | 43.46 | 61.75 | 1.83 | 2.79 | 2.53 |
| **Standard Deviation** | 10.04 | 19.80 | 0.92 | 1.28 | 1.19 |
| **Sample Size** | 115 | 115 | 116 | 117 | 119 |
| **Std. Error of mean** | 0.92 | 1.85 | 0.09 | 0.12 | 0.11 |
| **Minimum** | 20 | 16 | 1 | 1 | 1 |
| **Median** | 40 | 60 | 2 | 3 | 2 |
| **Maximum** | 90 | 140 | 5 | 5 | 5 |

Management

|  | PHB | Boss resists change | Boss impedes progress | Boss hrs. vs. methods | Boss short-term goals | Boss communicates |
|---|---|---|---|---|---|---|
| **Mean** | 2.08 | 2.46 | 2.35 | 2.26 | 2.70 | 3.30 |
| **Standard Deviation** | 1.25 | 1.24 | 1.27 | 1.26 | 1.41 | 1.25 |
| **Sample Size** | 113 | 67 | 116 | 114 | 115 | 116 |
| **Std. Error of mean** | 0.12 | 0.15 | 0.12 | 0.12 | 0.13 | 0.12 |
| **Minimum** | 1 | 1 | 1 | 1 | 1 | 1 |
| **Median** | 2 | 2 | 2 | 2 | 2 | 4 |
| **Maximum** | 5 | 5 | 5 | 5 | 5 | 5 |

Management (cont.)

| Boss skills needed | # | % |
|---|---|---|
| People | 30 | 26.32 |
| Project | 60 | 52.63 |
| Technical | 19 | 16.67 |
| Blank | 5 | 4.39 |
| Total | 114 | 100.00 |

| Boss bkgd | # | % |
|---|---|---|
| Business | 20 | 16.67 |
| Technical | 81 | 67.50 |
| Other | 13 | 10.83 |
| Don't know | 3 | 2.50 |
| N/A | 3 | 2.50 |
| Total | 120 | 100.00 |

Labor

| | Co-workers skills | Co-workers change | Co-workers communicate |
|---|---|---|---|
| Mean | 3.33 | 2.70 | 3.32 |
| Standard Deviation | 1.28 | 0.99 | 0.93 |
| Sample Size | 115.00 | 92.00 | 114.00 |
| Std. Error of mean | 0.12 | 0.10 | 0.09 |
| Minimum | 1 | 1 | 1 |
| Median | 4 | 3 | 3 |
| Maximum | 5 | 4 | 5 |

| Co-workers skills | # | % |
|---|---|---|
| people | 25 | 20.83 |
| project | 36 | 30.00 |
| technical | 54 | 45.00 |
| blank | 5 | 4.17 |
| total | 120.00 | 100.00 |

Workplace

| Space | Quiet |
|---|---|
| 3.81 | 3.51 |
| 1.09 | 1.20 |
| 91.00 | 90.00 |
| 0.11 | 0.13 |
| 1 | 1 |
| 4 | 4 |
| 5 | 5 |

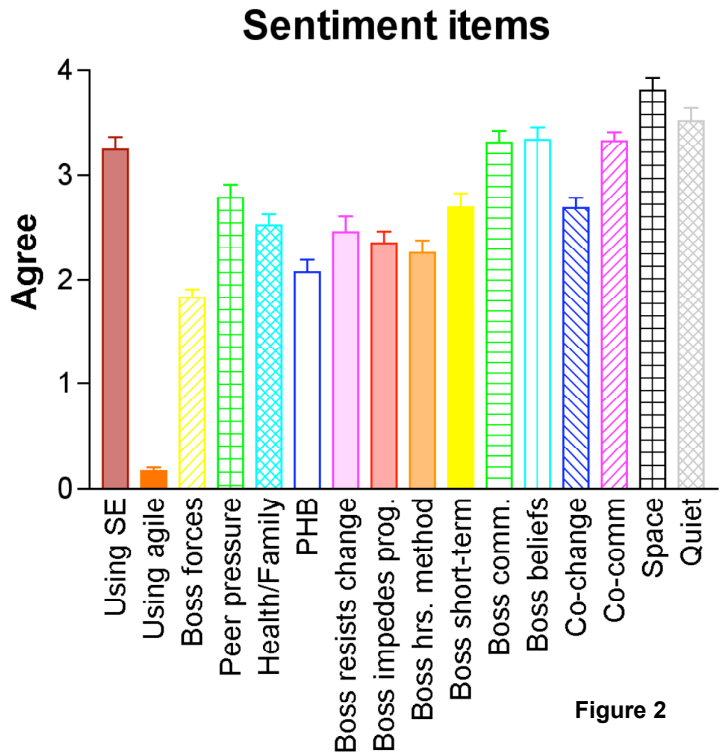

Figure 2

## ANOVAs

A one-way non-parametric ANOVA revealed significant differences between the business, other, and technical background groups. Significant mean differences of 1.589 and 1.495 were observed between the business group and the technical/other groups respectively with regard to "My supervisor is a PHB" sentiment (Fig. 3). A sig. mean difference of 1.117 ($p<.01$) was observed between the business and technical background groups with regard to the "short-term goals vs. long term productivity" item (Fig. 4). A sig. difference of 1.024 ($p<.01$) was observed between the business and technical groups with regard to a preference for longer hours over changes to methodology (Fig 5).

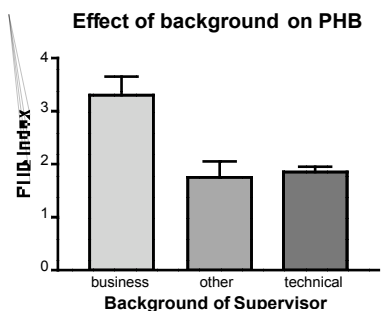
Figure 3

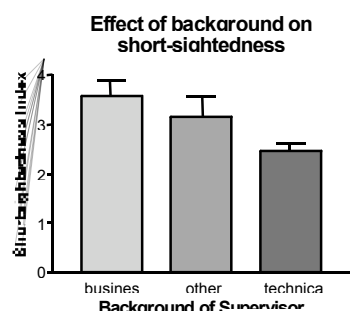
Figure 4

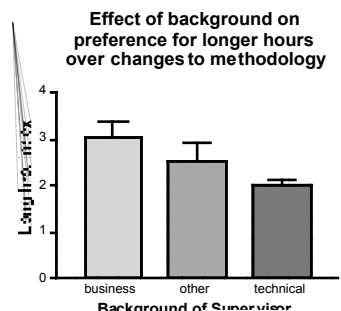
Figure 5

No connection between the background of supervisors and the level of software engineering in place or failure rate was detected. No significant difference was seen between the skills people wanted their supervisors to have and the sharing of beliefs about work.

**Background-skill matrix**

|  | Skill needed | | |
|---|---|---|---|
| **background** | people | technical | project |
| business | 4 / 20% | 6 / 30% | 10 / 50% |
| other | 1 / 9% | 3 / 27% | 7 / 64% |
| technical | 23 / 32% | 40 / 55% | 9 / 13% |

Chi-square analysis revealed background and skill variables were significantly associated ($p < .0003$)

**Correlation Matrix**

| | Age | # supervise | Using SE practices | Using agile methods | % failure | Avg. hrs. | Max hrs. |
|---|---|---|---|---|---|---|---|
| **Age** | 1 | | | | | | |
| **# supervise** | ns | 1 | | | | | |
| **Using SE practices** | ns | ns | 1 | | | | |
| **Using agile methods** | ns | ns | ns | 1 | | | |
| **% failure** | ns | ns | ns | ns | 1 | | |
| **Avg. hrs.** | ns | r=.1963 p=.0264 | ns | ns | ns | 1 | |
| **Max hrs.** | ns | ns | ns | ns | ns | r=.5403 p<.0001 | 1 |
| **Boss forces long hours** | ns | ns | ns | ns | r=.3559 p=.0003 | r=.3266 p=.0003 | ns |
| **Peer pressure** | ns | ns | ns | ns | ns | ns | ns |
| **Health & Family** | ns | r=.2059 p=.0247 | ns | ns | r=.3227 p=.0009 | r=.4297 p<.0001 | r=.3678 p<.0001 |
| **PHB** | ns | r=.2478 p=.0061 | r=-.2551 p=.0055 | ns | r=.2867 p=.0042 | ns | ns |
| **Boss resists change** | ns | ns | r=-.2881 p=.0210 | ns | ns | ns | ns |
| **Boss impedes progress** | ns | ns | r=-.3399 p<.0001 | ns | r=.2118 p=.0335 | ns | ns |
| **Boss hrs. vs. methods** | ns | ns | r=-.4341 p<.0001 | ns | r=.3280 p=.0009 | ns | ns |
| **Boss short-term goals** | ns | r=.1901 p=.0419 | r=-.2949 p=.0016 | ns | r=.2169 p=.0302 | ns | ns |
| **Boss comm..** | ns | r=-.2280 p=.0138 | r=.3646 p<.0001 | ns | ns | ns | ns |
| **Boss shares wrk. beliefs** | ns | ns | r=.4070 p<.0001 | ns | r=-.2174 p=.0290 | ns | ns |
| **Co-wrk resist change** | ns | ns | ns | ns | ns | ns | ns |
| **Co-wrk comm..** | ns | r=-.2073 p=.0220 | r=.1916 p=.0377 | ns | ns | ns | ns |
| **Space** | ns | ns | ns | ns | ns | ns | ns |
| **Quiet** | ns | ns | ns | ns | ns | ns | ns |

|  | Boss forces long hours | Peer pressure | Health & Family | PHB | Boss resists change | Boss impedes progress | Boss hrs. vs. methods |
| --- | --- | --- | --- | --- | --- | --- | --- |
| **Boss forces long hours** | 1 | | | | | | |
| **Peer pressure** | r=.3039 p=.0010 | 1 | | | | | |
| **Health & Family** | r=.4635 p<.0001 | r=.3459 p<.0001 | 1 | | | | |
| **PHB** | ns | ns | r=.2387 p=.0113 | 1 | | | |
| **Boss resists change** | ns | r=.3050 p=.0135 | ns | r=.4409 p=.0002 | 1 | | |
| **Boss impedes progress** | ns | ns | r=.1944 p=.0373 | r=.4889 p<.0001 | r=.6715 p<.0001 | 1 | |
| **Boss hrs. vs. methods** | r=.3136 p=.0008 | ns | r=.3037 p=.0011 | r=.4793 p<.0001 | r=.6448 p<.0001 | r=.6704 p<.0001 | 1 |
| **Boss short-term goals** | r=.2424 p=.0100 | ns | r=.3072 p=.0009 | r=.5198 p<.0001 | r=.6622 p<.0001 | r=.6529 p<.0001 | r=.6880 p<.0001 |
| **Boss comm..** | r=-.2135 p=.0232 | ns | ns | r=-.4530 p<.0001 | r=-.5073 p<.0001 | r=-.5424 p<.0001 | r=-.4688 p<.0001 |
| **Boss shares wrk. beliefs** | r=-.3074 p=.0010 | ns | r=-.1945 p=.0381 | r=-.5634 p<.0001 | r=-.5519 p<.0001 | r=-.6309 p<.0001 | r=-.5807 p<.0001 |
| **Co-wrk resist change** | ns | ns | ns | ns | ns | ns | r=.2368 p=.0264 |
| **Co-wrk comm..** | ns | ns | ns | ns | ns | ns | ns |
| **Space** | ns | r=.2826 p=.0076 | ns | r=-.3274 p=.0020 | ns | ns | ns |
| **Quiet** | ns | ns | ns | ns | ns | ns | ns |

|  | Boss short-term goals | Boss comm.. | Boss shares wrk. beliefs | Co-wrk resist change | Co-wrk comm.. | Space | U: Quiet |
| --- | --- | --- | --- | --- | --- | --- | --- |
| **Boss short-term goals** | 1 | | | | | | |
| **Boss comm..** | r=-.5566 p<.0001 | 1 | | | | | |
| **Boss shares wrk. beliefs** | r=-.6857 p<.0001 | r=.6650 p<.0001 | 1 | | | | |
| **Co-wrk resist change** | r=.2098 p=.0484 | ns | ns | 1 | | | |
| **Co-wrk comm..** | r=-.2030 p=.0326 | r=.3500 p=.0002 | r=.2102 p=.0275 | r=-.4142 p<.0001 | 1 | | |
| **Space** | ns | ns | r=.2797 p=.0087 | ns | ns | 1 | |
| **Quiet** | ns | r=.2397 p=.0253 | ns | ns | ns | r=.5853 p<.0001 | 1 |

# Discussion

**Supervisors**
Interestingly, respondents who supervised more people were more likely to cite communication with supervisors (r=-.2280, p=.0138) and coworkers (r=-.2073, p=.0220) as a problem.  They were also more likely to identify their supervisor as a PHB (r=.2478, p=.0061) and work longer hours on average (r=.1963, p=.0264).  The PHB correlation may be attributable to these individuals working closer to the executive strata.

**Software engineering principles**
Perhaps not surprising, the use of software engineering principles was negatively correlated with PHB's (r=-.2551, p=.0055) and related PHB characteristics, especially the tendency to favor longer hours over new methodologies (r=.4341, p<.0001).  The use of these principles was correlated with a sharing of work beliefs (r=.4070, p<.0001) and better communication with supervisor (r=.3646, p<.0001) and to lesser extent coworker (r=.1916, p=.0377).  In turn, communication with boss was correlated with that of the workers with each other (r=.3500, p=.0002)

**Failure**
Those who registered a high failure rate were more likely to feel their supervisor was extending their workweek (r=.3559, p=.0003) and more likely to experience health/family issues as a result of those long hours (r=.3227, p=.0009).  Ironically, their hours were not significantly longer than those in more successful situations, which indicates the experience of failed or futile death marches may have a greater impact than the hours themselves.

Those who experience failure were more likely to identify their supervisor as a PHB (r=.2867, p=.0042) with associated characteristics like impedance to progress (r=.2118, p=.0335), preoccupation with short-term goals (r=.2169, p=.0302) and a discordance of work beliefs (r=-.2174, p=.0290).

The definition of failure, poorly thought out before the survey was developed, became something of a sore point for respondents.  After some consideration, an option was given to allow survey takers to give their own definition of a failed project.  Responses included nonconformity to specifications, failure to be picked up by the customer, or failure in making a profit for the company.  Others were more reticent to view their projects as failures. "They only fail if you abandon them," wrote one respondent.  Success or failure is likely to be viewed economically and perhaps without any connection to software engineering.  A project can also fail at any point in time - from conception to maintenance.  All this begs the question of why the term failure is used so casually in the literature without any consensus over its definition.

**Hours & Pressure**
As expected, long workweeks were correlated with a higher report of health and family problems (r=.4297, p<.0001).  Although no correlation was observed between peer pressure and average hours, those who reported pressure from their boss and peers to work longer hours did not actually work significantly longer than others, but then reported higher incidence of health and family problems from these peer (r=.3459, p<.0001) and supervisor (r=.4635, p<.0001) pressure.  Once again indicating that the overall situation, more than absolute numbers, has a bigger effect on employee morale and job satisfaction.

Although not directly related to the industry, companies or management, this survey indicates peer pressure is a strong factor in driving up hours.  This pressure might be more significant in software engineering because of the need for putting in "face time" even when not a critical link in the development process.  Huge disparities in productivity among different programmers can lead to longer hours in an effort to match up.  A recent study by found that peer pressure and deadlines were a bigger factor in extending the hours of high-tech workers than supervisors. [SO02]

**Supervisor background**
No significant difference was found between the background groups with respect to the use of software engineering principles or failure rate.  However, significant differences were found in the areas of shortsightedness (Fig. 4), preference for longer hours over changes to methodologies (Fig. 5), and the PHB index (Fig. 3).  While managers from business backgrounds do show some bad qualities over their more technically inclined brethren, these differences were not across the board and did not show up on two critical items - adoption of software engineering principles and failure rate.

The author feels that a ANOVA with a slightly bigger sample size may yield significant differences on those endpoints because the business background was associated with PHB sentiment (Fig. 3), which in turn was correlated with both SE principles (r=-.2551, p=.0055) and failure rate (r=.2867, p=.0042).

**Skills**
Results of the background skills matrix proved to be intriguing.  Respondents who worked for business background supervisors were roughly four times more likely to recommend improved project management skills for their boss than those who worked for technical managers.  Ironically, employees indicated that technical managers were most likely to be in need of more technical skills (55%).  This may be attributable to the differing roles of PHB's and NB's in software development.  In retrospect, the phrasing of the items, "What my supervisors/co-workers need most are…" were somewhat ambiguous in retrospect – what someone needs doesn't necessarily imply that they don't already have it.

**Dead Ends**
No other variable was significantly correlated with age, agile methodologies, or workplace factors. This last item may have suffered from the decreased sample size of these items due to their addition to the survey halfway into its existence. Work hours can largely be regarded as a matter of personal preference, although one respondent indicated his or her most productive hours were "when no one else is here."

**Conclusion**
This confluence of PHB index and business-background managers together with the correlation of PHB index and failure rate/SE adoption might raise some doubts concerning the suitability of business-background managers to direct software engineering projects, at least at the programmer level. However, the skills matrix indicates that this type of management is most likely in need of project management skills, not technical skills.

Perception of one's work viability and pressure from others proved significant in many endpoints originally thought to be solely correlated with raw workload, such as health and family detriment.

Although perhaps obvious from an anecdotal perspective, the importance of motivation should not be underestimate in the contexts of these results. Motivation can come from working on an inherently exciting or well-managed project, but managers must motivate work with their developers by assigning meaningful and challenging work, rewarding excellence, coaching, and "convincing [employees] that their job is vital for the business to succeed." [MM00]

## Future Directions
A survey with a larger, more controlled audience would have undoubtedly yielded even more significant results. The area of ethics, which the author does not feel would be adequately surveyed in such an uncontrolled method, is one topic that could be investigated in such a study. Other areas that might tie into to this study are traditional leadership issues of motivation and trust. A survey composed solely of managers would be useful to explore the same issues discussed in this article from their perspective.

In terms of solutions to the area of death marches, one can focus on the industry, company practices, and management. Management behaviors might be the easiest of these three to change, if only because developers can exercise some power over management through constructive criticism and, of course, migration.

While the author does not presume to know how to change peoples work behaviors, there may be is some value in encouraging developers and managers to see work situations from each others point of view. The key to dealing with many interpersonal problems is to try to understand the values, behaviors, and beliefs of the cultural groups involved in a project. [MK98]